\documentclass[twocolumn,showpacs,preprintnumbers,amsmath,amssymb]{revtex4}
\usepackage{graphicx}
\usepackage{dcolumn}
\usepackage{bm}
\usepackage{amssymb}
\usepackage{amsmath}
\begin{document}
\title{Non linear transport properties of an insulating YBCO nano--bridge}

\author{L. Fruchter, A. Kazumov, J. Briatico, A.A. Ivanov, V. Nicholaichik}%
\affiliation{Laboratoire de Physique des Solides, C.N.R.S. Universit\'{e} Paris-Sud, 91405 Orsay cedex, France \\
UMR-CNRS/THALES, Route D128, 91767 Palaiseau, France\\
Institute of Microelectronics Technology, Chernogolovka, 141432 Moscow Region Russia\\
Moscow Engineering Physics Institute, Kashirskoe sh. 31, 115409 Moscow Russia
}
\date{Received: date / Revised version: date}
%
\begin{abstract}{}
We have investigated the transport properties of an insulating sub-micrometric YBa$_2$Cu$_3$O$_{7-\delta}$ bridge, patterned on a thin film. As expected for a variable-range-hopping insulator, transport is found non linear. The reduced dimension allows for the observation of an individual fluctuator generating random telegraph noise, which dynamics could be explored as a function of the temperature and transport current. Some recordings clearly exhibit three levels fluctuating resistance, with comparable level separation and correlated dynamics, which likely result from the existence of two states or correlated clustered charge traps.
\end{abstract}


\maketitle
%
It is well known that superconductivity in cuprates emerge from doping a Mott insulator to obtain the underdoped superconductor\cite{Lee2004}. As such, as pointed out early, high-Tc superconductors share similarities with doped two-dimensional weakly compensated semiconductors\cite{Edwards1995,Edwards1998}. In the insulator, carriers are localized within a localization length (effective Bohr radius) which diverges - together with dielectric constant - at the transition to the delocalized state, at some critical value of the tuning parameter, $x=x_c$, where $x$ may be a carrier density, disorder strength, impurity concentration, magnetic field... In the perspective of a quantum phase transition between adjacent insulating and superconducting states, the variable range hopping (VRH) law for the insulator conductivity may be viewed as one example of a scaling law associated to a quantum critical point, in which the divergence of the localization length at the critical density for doping accounts for the critical temperature scaling of the form $\rho(T, n) = F(\delta/T^{1/z\nu})$, where $\delta = (x-x_c)/x_c$, $z$ is the dynamical exponent and $\nu$ is the correlation length exponent. In the regime of hopping in the Coulomb gap\cite{Efros1975}, ones has $z=1$, as expected for a strongly interacting Coulomb system\cite{Sondhi1997}. There are several experimental indications for the classical metal--insulator transition that the VRH regime might be characteristic of the intrinsic insulating phase, representing the disordered state to be found on one side of such a quantum transition, even in the absence of strong Coulomb interaction\cite{Hamilton2001,Parendo2006}. In the case of strongly disordered systems, the insulating phase could be found to coexist with the metallic or superconducting one\cite{Hamilton2001,Fruchter2008} and may influence the transport properties far into the metallic phase\cite{Fruchter2007}.
Besides classical scaling dependence for the conductivity, the insulating phase is also characterized by the existence of slow dynamics for charges localized on traps. In the modelization by a percolation resistance network, such a behavior arises from the existence of a network of thermally assisted hopping paths for conduction, and neighboring sites loosely coupled to this network\cite{Shklovskii1980,kozub1996}. Due to the exponential dependence of the hopping rate on site distance, there is a wide distribution of hopping rates from such sites, which accounts for a $1/f$ resistance noise\cite{Shklovskii2003}. Reducing the dimension of the conductor may allow to resolve some individual contribution of the fluctuators to the resistance. In particular, sites with occupancy varying slowly on the time scale of the experiment may be evidenced by random telegraph noise.

\begin{figure}
\resizebox{0.7\columnwidth}{!}{\includegraphics{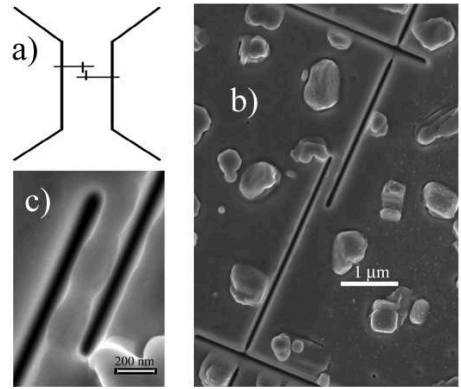}}
\caption{a) Schematics for the FIB process from a lithographied film, b) actual realization on a YBa$_2$Cu$_3$O$_{7-\delta}$ film, c) detail of the bridge.}\label{sketch}
\end{figure}

We have investigated the transport properties of such a conductor, from a sub-micron bridge patterned on YBa$_2$Cu$_3$O$_{7-\delta}$ thin film. There are indications that cuprates high-Tc superconductors may provide one example of a quantum critical point, at the superconductor--insulating transition\cite{Hetel2007}. For weakly disordered compounds, such as the present one, there should be no mixing of the two phases, nor existence of some intermediate glassy phase at the transition. It is however, notoriously difficult to prepare such bridges, with a controlled doping state. The reason for this is likely the ability of the doping oxygen to migrate from narrow bridges, yielding patterned samples that are less doped than the starting material. Complex preparation of the bridges, combining appropriate deposition of protective layers and reactive ion etching, such as in ref.~\cite{Bonetti2004}, could provide bridges with unchanged doping state. However, more direct techniques, such as focused ion beam (FIB) etching, likely induce a strong reduction of doping, so that the observation of superconductor--insulator transition with the reduction of the bridge dimension may actually be due to the concomittent reduction in the doping state\cite{Mikheenko2005}. The migration of the oxygen out of the material could be due in this case to both the increased local temperature during etching, and to electrical field effects created by the charged ions deposited by FIB. The film was grown on a SrTiO$_3$ substrate, using laser ablation\cite{Ivanov1991}, and pre--patterned using optical lithography. The as-grown sample displayed sharp superconducting transition (0.5~K wide) at $T = 92$~K; the thickness was $4000$~\AA. As very often observed for thick YBa$_2$Cu$_3$O$_{7-\delta}$ films, electron microscopy revealed the existence of well defined large inclusions, emerging from the flat surface of the film. Such inclusions imposed to increase the Ga ions FIB dose to cut the sample along the set of two long lines leading to the bridge itself, but they were not a problem for the bridge itself, which was patterned using two short cuts intersecting the long ones, in some selected area of the film free from any inclusion. The short cuts defining the bridge were made from one hundred sweeps of the FIB, etching the film down to the substrate (Fig.~\ref{sketch}). While the geometrical size for the bridge was about $0.2$ x $l = 0.8\,\mu$m$^2$, the actual conducting width was likely much smaller, due to the material degradation along the FIB lines. The corresponding upper limit for the resistivity was $\rho(300K) < 1\,\Omega$~cm. 

\begin{figure}
\resizebox{0.7\columnwidth}{!}{\includegraphics{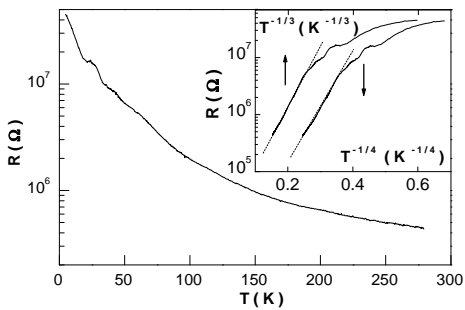}}
\caption{Resistance of the constriction ($I = 1$ nA), in representations where a VRH would appear linear (2D : $T^{-1/3}$; 3D : $T^{-1/4}$). Both VRH laws fail below $T\approx 45$~K.}\label{resistance}
\end{figure}

\begin{figure}
\resizebox{0.7\columnwidth}{!}{\includegraphics{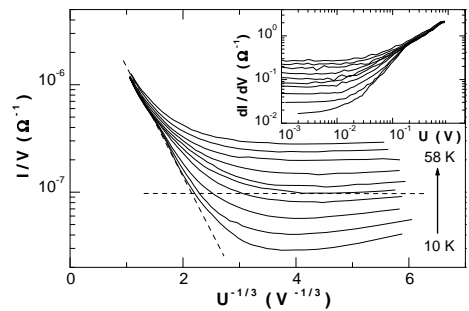}}
\caption{Non linear conductance, in a representation where a 2D VRH gives a straight line at low electric field and zero temperature. From bottom to top : $T = 10, 14, 18, 26, 29, 34, 40, 46, 50, 58$ K. Inset : the same data, in a representation similar to the one used in ref.~\cite{Christiansen2002}. The dashed lines show the determination of a characteristic electric field at $T =$ 29 K.}\label{VE}
\end{figure}

Starting from the optimally doped film, the resulting bridge was strongly insulating. Cycling the sample temperature from the ambient down to 4.2 K resulted in a systematic increase of the resistance, and the results shown below are for one of such cycles. As can be seen in Fig.~\ref{resistance}, one or the other of the 2D and 3D VRH law (respectively $R \propto$ exp$(T_0/T)^{1/3}$ and $R \propto$ exp$(T_0/T)^{1/4}$) may describe the data above $T \simeq 45$~K, but the resistance is systematically smaller than expected for lower temperature (in what follows, we will adopt a two-dimensional description, owing to the increased anisotropy for YBCO at low doping\cite{Janossy1991}, however stressing that both behaviors are indiscernible from the present data). The characteristic temperature in this case is $T_0 \simeq 1.5\,10^4$~K from the data in Fig.~\ref{resistance}). 
Actually, the measured resistivity was strongly non-linear at low temperature (Fig.~\ref{VE}) and this might explain the bending of the data in Fig.~\ref{resistance}. It is indeed well known that the effect of an electric field in the VRH regime is to increase the effective temperature of the carriers (which determines the site occupation, using a Fermi distribution at the corresponding temperature), crossing from $T_{eff} = T$ at low field to $k_B T_{eff}\simeq e E a$ at high field, where $a$ is the localization length\cite{Marianer1992}. Our data is similar to the one in ref.~ \cite{Kinkhabwala2006}, where a complete numerical computation of the effect of the electric field was given. An estimate for the localization length may be obtained, using for the characteristic electric field at $T = $ 29~K, $E(T) = U/l \simeq 1.2\,10^5$~Vm$^{-1}$ (Fig.\ref{VE}), and $E_c \simeq k_B T / e a$ (\cite{Shklovskii1973,Kinkhabwala2006}), yielding $a \simeq 200$ \AA. We remark that the differential resistance may appear to present a threshold voltage (Fig.~\ref{VE}, inset), as proposed in ref.~\cite{Christiansen2002}. In our case, this is a consequence of the non linearity for electrical field $E \approx E_0$. In particular, we rule out the Coulomb blockade as the origin of the non--linearity, as a tentative threshold voltage, $U \simeq 0.2$~V (Fig.\ref{VE}), would correspond to Coulomb energies $\simeq 10^3$~K, whereas the I-V characteristic strongly evolve at much lower temperature.

Non linear transport phenomena must be systematically confronted to a possible contribution of Joule heating, and one should check whether the actual temperature of the thin film sample could be here significantly different from the one of the substrate. First, we shall see below that the localization length derived from non linearity agrees with the one derived from the dynamics of a single fluctuator, whereas Joule heating could potentially provide just any value for $a$. Also, fluctuator switching events are found thermally activated down to the lower temperature explored, whereas a saturation would be expected for significant heating of the sample. Finally, we may estimate the overheating, in  a way similar to what was done in ref.~\cite{Lavrov2003}. The temperature of the bridge may be evaluated from $T_{br} \simeq T+ 2E^2S[\rho(T_{br}) \tilde{\kappa}]^{-1}$, where $T_{br}$ is the bridge temperature, $S$ its cross section, $\rho$ the material resistivity and $\tilde{\kappa}$ an effective heat conductivity for the substrate. Using the resistance of the bridge, the heating is actually independent of the effective cross section. The high resistance helps limiting the temperature increase in our case. Solving this equation, using the temperature independent conductivity $\tilde{\kappa} =$ 150 mW/K cm and the resistance data in Fig.~\ref{resistance}, we find that the heating does not exceed 1 K at 30 K for the higher electric field of our data.

\begin{figure}
\resizebox{0.7\columnwidth}{!}{\includegraphics{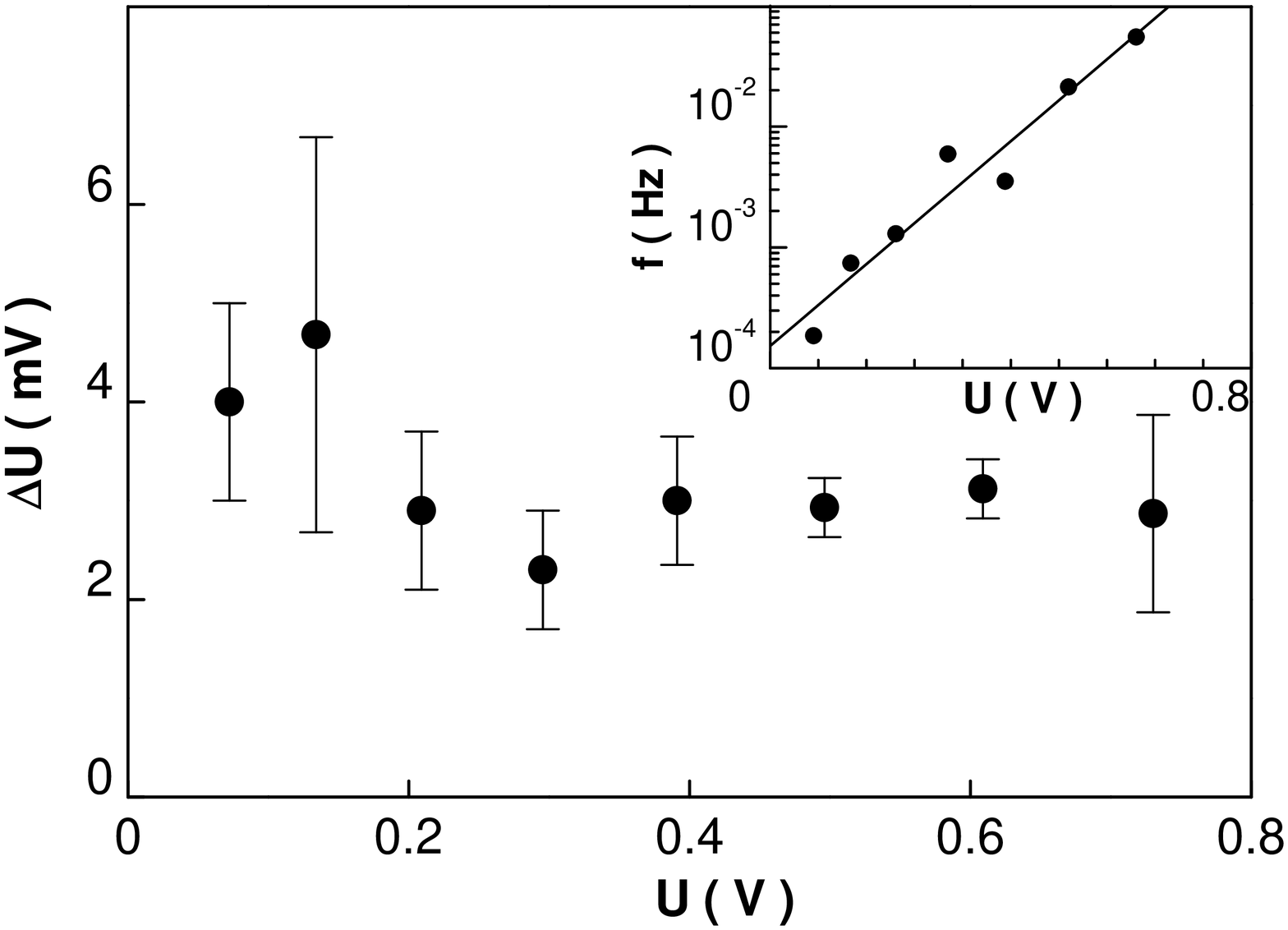}}
\caption{Mean voltage jump for different driving voltage. Inset: switching rate, showing thermally activated behavior. The line is a linear fit, yielding $a = 170$ \AA ($T = 25$ K.). Note that the switching rate becomes unmeasurably small upon entering the linear regime displayed in Fig.~\ref{VE}.} \label{dynamicsT25}
\end{figure}

\begin{figure}
\resizebox{0.7\columnwidth}{!}{\includegraphics{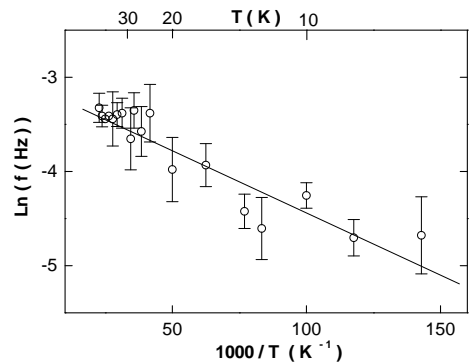}}
\caption{Arrhenius plot of the switching frequency ($I = 1\, \mu$A). The line is the best fit to thermally activated behavior, with activation energy 13 K.} \label{fvsT}
\end{figure}

The recording of the voltage during a temperature sweep in the non linear regime revealed clear jumps, between two discrete values, in a telegraph--like fashion. Voltage time recording, at a fixed temperature and driving current, confirmed the existence of a voltage telegraphic noise. The size of the voltage jumps were essentially temperature independent. Also, the voltage jumps magnitude was roughly independent from the driving current (Fig.~\ref{dynamicsT25}). However, the switching rate between descrete states was strongly dependent upon the voltage (alt. the current) applied to the constriction, larger voltage yielding larger switching rate (Fig.~\ref{dynamicsT25}). Within experimental uncertainties (which essentially arise from the difficulty to extract the telegraphic noise from the measurement noise), the switching rate is found thermally activated at all temperature investigated (Fig.~\ref{fvsT}). We note that the base temperature, rather than $T_{eff}$, should enter the Arrhenius plot for the determination of the activation energy, as the trapped carriers do not experience the electric field acceleration. On the other hand, the observation of a strong influence of the voltage on the switching frequency implies that the electric field work, over a distance of the order $a$, lowers some activation energy for trapping, which is then expected of the form $E_0 - e a E$. Using the data in Fig. \ref{dynamicsT25}, we obtain from the slope of the linear fit of the inset, $dLn(f)/dU = e\,a /(l\,k_B T)$, $a =$ 170 \AA~ and $E_0/k_B =$ 220 K, the localization length being in full agreement with our former independent determination.
It should be stressed that we did not observe any switching event in the linear regime evidenced in Fig.~\ref{VE}: if such switching events exist in the linear regime, the rate must be so low for the currents involved for this regime, that it is out of our experimental time scale. As a consequence, we believe that the telegraphic noise observed here is characteristic of the non linear regime, and cannot easily be interpreted using standard linear models for random telegraph noise\cite{Kirton1989}, although the origin is likely the same : the capture of carriers into localized defects which either directly remove carriers from the conduction band, or alter the mobility of the carriers. Many such events, in larger samples, may be at the origin for $1/f$ noise in the vicinity of the superconductor--insulating transition\cite{Fruchter2008}.

\begin{figure}
\resizebox{0.7\columnwidth}{!}{\includegraphics{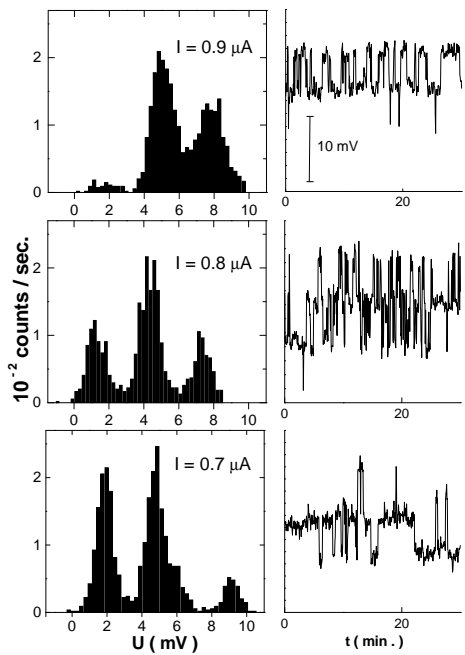}}
\caption{Left : three states voltage distribution for driving current $I=0.7, 0.8, 0.9\,\mu$A (the origin for $U$ was arbitrarily shifted). The weight of the distribution is progressively transfered to the upper level with increasing current. Right : part of the time recording used to establish the distribution.}\label{discurrent}
\end{figure}

For some recordings, the signal exhibited three states histograms, as shown in Fig.~\ref{discurrent}. The simplest interpretation is that such a signal results from the presence of two distinct traps, which independently capture an release charges. However, we systematically observed that the signal level separation was nearly the same between adjacent levels. This could have been only fortitious, but we have observed that the levels occupancy evolved with driving current in a correlated manner: for independent traps, such an increase should simply increase the switching rates of both traps, whereas we observe that the weight of the distribution is progressively transfered from the lower level to the upper one (Fig.~\ref{discurrent}). This correlated behavior suggests that, rather than two independent traps, one actually has two traps correlated in some way, or a single trap with three states. 

Such a case was first reported for the linear conductance of a MOSFET channel\cite{Uren1988}. As underlined by the authors, the correlated two traps case requires that these should reside at a close distance, so that Coulombic interaction between them can shift the energy level of the trapped charge by several $k_BT$ : this implies clustering of the defects at point--like nucleation sites, at a distance that does not, however, allow for hybridization of the defects. The second possibility is the existence of traps which can accommodate multiple charges, as reported in MOSFETs\cite{Uren1988,Nakamura1989}. The nature of such three states traps is however unclear. In the case of Anderson-localized states, the possibility for doubly--occupied localized states exists as, when the on--site correlation energy, $U$, is smaller than the bandwidth of the localized states distribution, some of the localized sites may be doubly occupied, with two states\cite{Kamimura1980}. The upper state energy is close to $E_F$, as for singly occupied states, while the lower state energy is about $U$ below the Fermi level. However, for the higher electrical field reached in the present study, $E/E_0 \simeq 0.1$, we have $T_{eff}/T_0 \simeq 0.1 \simeq 10^3$ K, which is too low to allow for the activation of the state lying typically\cite{Kastner1998} $U \simeq 2$ eV $\simeq 2\,10^4$ K below $E_F$. Then, transport in our moderate out--of--equilibrium conditions should involve only the higher states, as is the case for the linear transport regime\cite{Kurobe1982}. Finally, one should also consider the possibility for a contribution of collective charge motion. Indeed, it has been proposed that charges may gather into conductive channels -- or "`stripes"' -- as a Mott insulator is doped. As underlined in Ref.~\cite{Lavrov2003}, collective motion of charges generally implies a sharp threshold electric field and a narrow band noise. The latter property should however be characteristic of conventional charge order extending along large coherence length, rather than of quasi--1D stripes, as expected in cuprates, so that what one really would expect, in the present case, are a low threshold electric field and large conductance fluctuations, associated to the existence of a large number of correlated charges. These two characteristics were both observed in the present study, but we have favored a more conventional interpretation based on transport in the VRH regime. We find that it is able to account for several of our observations but, clearly, the exact mechanism for some correlation that we have observed for the trapped charges needs to be investigated in a more systematic way. While our tuning parameters for the dynamics of an individual fluctuator was restricted here to the voltage that tunes the accessible states in the non--linear VRH regime, and the temperature that allows for the relaxation of the charges from traps, it would be desirable to be able to investigate the behavior of individual fluctuators in the linear regime also. Reversibly tuning the doping of the material over some short range, using the field effect on very thin films, just as one tunes the channel conductivity of a MOSFET from the gate voltage, appears as a convenient way to do so.

\scriptsize{V.I.Nikolaichik and A.A.Ivanov acknowledge the support of the ISTC (Moscow) under the Project No.3357 and the RFBR under the Project No.08-02-00759a. L. Fruchter and J. Briatico acknowledge the support of the A.N.R. under project No.ANR-07--1--19--3024. The authors are grateful to F. Fortuna from Institut d'Electronique Fondamentale for F.I.B. technical support.}
%

\begin{thebibliography}{}

\bibitem{Lee2004}P.A. Lee, Naoto Nagaosa, Xiao-Gang Wen, Rev. Mod. Phys. \textbf{78}, 17 (2004)
\bibitem{Edwards1995}Edwards P.P., Ramakrishnan T.V. and Rao C.N.R., J. Phys. Chem. \textbf{99}, 5228(1995)
\bibitem{Edwards1998}Edwards P.P., Mott N.F. and Alexandrov A.S., Journal of Superconductivity \textbf{11}, 151 (1998)
\bibitem{Efros1975}A. L. Efros and B. I. Shklovskii, J. Phys. C \textbf{8}, L49 (1975)
\bibitem{Sondhi1997}S. L. Sondhi, S. M. Girvin, J. P. Carini and D. Shahar, Rev. Mod. Phys. \textbf{69}, 315 (1997)
\bibitem{Hamilton2001}Hamilton A.R., Simmons M.Y., Pepper M., Linfield E.H. and Ritchie D.A., Physica B \textbf{296}, 21 (2001)
\bibitem{Parendo2006}Parendo Kevin A., Sarwa K. H., Tan B. and Goldman A. M., Phys. Rev. B \textbf{73}, 174527 (2006)
\bibitem{Fruchter2008}L. Fruchter, Z. Z. Li and H. Raffy, Eur. Phys. J. B \textbf{65}, 213 (2008)
\bibitem{Fruchter2007}L. Fruchter, H. Raffy and Z.Z. Li, Phys. Rev. B \textbf{76}, 212503 (2007)
\bibitem{Shklovskii1980}B. I. Shklovskii, Solid State Com. \textbf{33}, 273 (1980)
\bibitem{kozub1996}V.I. Kozub, Solid State Com., \textbf{97}, 843 (1996)
\bibitem{Shklovskii2003}B. I. Shklovskii, Phys. rev. B \textbf{67}, 045201 (2003)
\bibitem{Hetel2007}I. Hetel, T. R. Lemberger and M. Randeria, Nature Physics \textbf{3}, 700 (2007)
\bibitem{Bonetti2004}J. A. Bonetti, D. S. Caplan, D. J. Van Harlingen, and M. B.Weissman, Phys. Rev. Lett. \textbf{93}, 087002 (2004)
\bibitem{Mikheenko2005}P. Mikheenko, X. Deng, S. Gildert, M. S. Colclough, R. A. Smith, C. M. Muirhead, P. D. Prewett and J. Teng,Phys. Rev. B \textbf{72}, 174506 (2005)
\bibitem{Ivanov1991}A.A. Ivanov, S.G. Galkin, A.V. Kuznetsov and A.P. Menushenkov, Physica C, \textbf{180}, 69 (1991).
\bibitem{Christiansen2002}C. Christiansen, L. M. Hernandez and A. M. Goldman, Phys. Rev. Lett. \textbf{88}, 037004 (2002)
\bibitem{Janossy1991}B. Janossy, D. Prost, S. Pekker and L. Fruchter, Physica C \textbf{181}, 51 (1991)
\bibitem{Marianer1992}S. Marianer and B. I. Shklovskii, Phys. Rev. B \textbf{46}, 13100 (1992)
\bibitem{Kinkhabwala2006}Yusuf A Kinkhabwala, Viktor A Sverdlov, Alexander N Korotkov and Konstantin K Likharev, J. Phys. Condens. Matter \textbf{18}, 1999 (2006)
\bibitem{Shklovskii1973}B. I. Shklovskii, Sov. Phys. -- semicond. \textbf{6}, 1964 (1973)
\bibitem{Lavrov2003}A. N. Lavrov, I. Tsukada and Yoichi Ando, Phys. Rev. B \textbf{68}, 094506 (2003)
\bibitem{Kirton1989}M. J. Kirton and M. J. Uren, Advances in Physics \textbf{38}, 367 (1989)
\bibitem{Uren1988}M. J. Uren, M. J. Kirton, and S. Collins, Phys. Rev. B \textbf{37} 8346 (1988)
\bibitem{Nakamura1989}H. Nakamura, N. Yasuda, K. Taniguchi, C. Hamaguchi and A. Toriumi, Jap. Journ. Appl. Phys. \textbf{28}, L2057 (1989)
\bibitem{Kamimura1980}Hiroshi Kamimura, Phil. Mag. \textbf{42}, 763 (1980)
\bibitem{Kastner1998}M. A. Kastner, R. J. Birgeneau, Shirane and Y. Endoh, Rev. Mod. Phys. \textbf{70}, 897 (1998)
\bibitem{Kurobe1982}A. Kurobe and H. Kamimura, Journ. Phys. Soc. Jap. \textbf{51}, 1904 (1992)

\end{thebibliography}
%

\end{document}